\documentstyle[aps,multicol,epsf]{revtex}
\begin{document}
\draft

\title{Invasion percolation and global optimization}

\author{Albert-L\'aszl\'o Barab\'asi$^*$} 
\address{IBM, T.~J. Watson
Research Center, P.O. Box 218, Yorktown Heights, NY 10598, USA}

\date{\today}

\maketitle

\begin{abstract}
Invasion bond percolation (IBP) is mapped exactly into Prim's
algorithm
for finding the shortest spanning tree of a weighted random graph.
Exploring this mapping, which is valid for arbitrary dimensions and
lattices, we introduce a new IBP model  that belongs to the same
universality class as IBP and generates the minimal energy tree
spanning the IBP cluster.
\end{abstract}

\pacs{??}

\begin{multicols}{2}

\narrowtext

Flow in porous medium, a problem with important practical
applications, has motivated a large number of theoretical and
experimental studies \cite{wong}.  Aiming to understand the complex
interplay
between the dynamics of flow processes and randomness
characterizing
the porous medium, a number of models have been introduced that
capture different aspects of various experimental situations.  One
of
the most investigated models in this respect is invasion
percolation
 \cite{Wilkinson83}, that describes low flow rate drainage
experiments or secondary migration of oil during the formation of
underground oil reservoirs \cite{wong,Vicsek92}.

When a wetting fluid (e.g. water) is injected {\it slowly} into a
porous medium saturated with a non-wetting fluid (e.g. oil),
capillary forces, inversely proportional to the local pore
diameter,
are the major driving forces determining the motion of the fluid. 
The
invasion bond percolation (IBP)
model captures the basic features of this invasion process.
Consider a two dimensional square lattice and assign random numbers
$p_{ij} \in [0,1]$ to bonds connecting the nearest neighbor
vertices
$x_i$ and $x_j$. Here $p_{ij}$ mimic the randomness of the porous
medium, corresponding to the random diameter of the pores, and
vertices correspond to throats. Invasion bond percolation without trapping is
defined
by the following steps: (i) Choose a vertex on the lattice. (ii)
Find
the bond with the smallest $p_{ij}$ connected to the occupied
vertex
and occupy it.  At this point the IBP cluster has two vertices and
one
bond.  (iii) In any subsequent step find the empty bond with the
smallest $p_{ij}$ connected to the occupied vertices, and occupy
the
bond and the vertex connected to it. 

The various versions of the model are
useful in matching the simulated dynamics  to the microscopic
effects acting as fluids with different wetting properties and
compressibility are considered.  Originally introduced to model
fluid
flow, lately invasion percolation is viewed as a key model in
statistical mechanics, investigated for advancing our understanding
of
irreversible and nonequilibrium growth processes with generic
scaling
properties \cite{Vicsek92}.

Finding the shortest spanning tree of a weighted random graph is a
well known problem in graph theory \cite{Chris75}.  Consider a
connected nondirected graph $G$ of $n$ vertices and $m$ bonds
(links
connecting vertices), with costs $p_{ij}$ associated with every
bond
$(x_i,x_j)$.  A {\it spanning tree} on this graph is a connected
graph
of $n$ vertices and $n-1$ bonds.  Of the many possible spanning
trees
one wants to find the one for which the sum of the weights $p_{ij}$
is
the smallest.  A well known example is designing a network that
connects
$n$ cities with direct city-to-city links (whose length is
$p_{ij}$)
and shortest possible total length.  This is a problem of major
interest in the planning of large scale communication networks and
is
one of the few problems in graph theory that can be considered
completely solved. Since for a fully connected graph with $n$
vertices
there are $n^{n-2}$ spanning trees \cite{Cayley1874}, designing an
algorithm that finds the shortest one in non-exponential time steps
is a
formidable  global optimization problem.

 An efficient algorithm for finding
the shortest spanning tree of an arbitrary connected graph $G$ was
introduced by Prim \cite{Prim57}, and involves the following steps:
(i) Choose an arbitrary vertex, $x_i$.  (ii) Of all vertices
connected
to $x_i$ find the one for which $p_{ij}$ is the smallest, and join
$x_i$ and $x_j$. (iii) At any subsequent step a new vertex is
appended
to the tree by searching for the bond that has the smallest weight
$p_{ik}$, where $x_i$ belongs to the tree, and $x_k$ does not. 
Thus
bonds that connect already occupied vertices are not eligible for
growth. 
It has been shown by Prim that the tree generated by the previous
algorithm is the smallest energy spanning tree for the graph $G$
\cite{proof}. Already at this point one can notice the formal
similarity between Prim's algorithm and the IBP model discussed
above.

In this paper I show the equivalence of the IBP model with Prim's
algorithm for finding the shortest spanning tree of a weighted
random
graph \cite{Prim57}, and explore the consequences of this
interesting mapping.  For this I introduce an invasion bond
percolation model with a local trapping rule (hereafter called IBPO
model).  At {\it every time step} the bonds invaded by the  IBPO
model
form the {\it minimum energy tree} spanning all vertices of the IBP
cluster, where energy is defined as the sum of the invaded random
bonds.  Moreover, the clusters generated by the IBPO model have the
same scaling and dynamic properties as the clusters of the standard
IBP
model.  Thus the two models (and Prim's algorithm) belong to the
same
universality class.  Since the IBPO cluster forms a tree (i.e.  is
loopless), this  result implies that loopless IBP belongs to the
same
universality class as  IBP. The cluster formed by the invaded
bonds coincides with the unique solution of the global optimization
problem of finding the smallest energy branching self-avoiding
walk  connecting all vertices of a finite lattice.  Furthermore,
the
IBPO model is computationally more efficient than the IBP model.
The
above results are {\it exact} and are valid for {\it arbitrary
dimensions} and {\it lattices}.

        The difference between the IBP and IBPO models comes in an
additional trapping rule \cite{trap}: in the IBPO model only bonds
connecting vertices of the cluster to empty vertices are eligible
for growth (see Fig. \ref{fig1}).  Note that in the IBP model there
may
be bonds eligible for growth, that connect two already occupied
vertices (hereafter these are be called {\it trapped} bonds, since
an
empty bond is trapped between two occupied  vertices).  In the IBPO
model these trapped bonds are not eligible for growth \cite{lung}.

        Consider the invasion process described by the  IBPO model,
and
assume that invasion ends when {\it all vertices} of a {\it finite
}
lattice have been invaded \cite{note-bonds}. The energy of the
obtained IBPO cluster is defined by $E=\sum p_{ij}$, where the sum
goes
over all {\it occupied} bonds.

 With these definitions  one can prove the following:

(a) The cluster generated by the IBPO model has the smallest energy
of
all possible clusters that span all vertices of the lattice.

(b) The obtained cluster is independent of the site chosen as the
starting point of the invasion process.

(c) Defining time as the number of invaded {\it vertices}, at any
time
step the vertices invaded by the IBPO model coincide with those
invaded
by the IBP model, implying that the IBPO and IBP models belong to
the
same universality class.

(d) At any time step the bonds invaded by the IBPO model form the
smallest energy tree spanning the vertices of the IBP cluster.

(e) The statements (a)-(d) are valid in any dimension and are
independent of the lattice.

In the following we discuss (a)-(e)  separately.

{\it (a)  Prim's algorithm and IBPO---}
Comparing the definition of the IBPO model and Prim's algorithm, we
find that  Prim's algorithm is {\it exactly} the IBPO model acting
on the graph
$G$.  Since the square lattice, for which the IBPO model is
defined, is
a particular case of an arbitrary graph, the cluster generated by
the
IBPO model coincides with the smallest energy spanning tree.

{\it (b)  Uniqueness of the IBPO cluster---} If  the IBPO
model selects the smallest energy tree, there is only one such a
tree,
provided that the $p_{ij}$'s are distinct real numbers, since the
chance of having two trees with the same number of bonds
and the same  energy is zero \cite{unique}.
Thus starting from any vertex of the lattice one should obtain the
same minimum energy cluster.

{\it (c)  Cluster properties---} A $B$-cluster is the set of {\it
bonds}
occupied
by the invasion process.  Similarly, a $V$-cluster is the set of
occupied {\it vertices}.  In percolation and fluid flow one is
interested in
the scaling properties of the first spanning cluster generated by
the
invasion algorithm.  In particular, it is known that clusters
generated by the IBP model are fractal. However, the fractal
dimension,
and in general the scaling exponents, may depend on the trapping
rule,
thus one needs to establish the universality class to which the
IBPO
model belongs, since it differs from the IBP model in a trapping
rule.

Defining time as the number of occupied {\it vertices}, at any time
step the $V$-clusters generated by the IBP and IBPO models are
identical \cite{vertex}, the only difference being that within one
time step
the IBP model may occupy a number of trapped bonds without adding
any
new vertex to the cluster. The IBPO model with every occupied bond
occupies a vertex as well.  In conclusion, at any time step the
{\it
$V$-clusters generated by the two models coincide}, provided, that
we
start the invasion process from the same vertex.  This implies that
the IBP and IBPO belong to the same universality class, and the
generated clusters have the same fractal dimension, whose value
coincides with the fractal dimension of ordinary percolation
\cite{univ}.

However, not only the static properties, but all dynamic properties
measured in terms of  the occupied  vertices
coincide as well. For example, the two models  generate exactly
the same set of avalanches \cite{Suki94,Fruenberg} and the growth
of
the cluster obeys the same dynamic scaling form \cite{Fruenberg}.

{\it (d)  Spanning trees and loopless percolation---}
Next I investigate the relation between the $B$-clusters generated
by
the two models.  The bonds invaded by the IBPO model is a subset of
the
bonds invaded by the IBP model, i.e.  at any time step $N^b_{IBP}
\ge
N^b_{IBPO}$, where $N^b_{IBPO}$ and $N^b_{IBP}$ are the number of
bonds
occupied by the IBPO and IBP models, respectively.  According to
(a)
and Prim's theorem, the bonds invaded by the IBPO model form the
smallest energy spanning path connecting the selected vertices.
Since
the IBP and IBPO models share the same vertices, at {\it  every
time step the
IBPO $B$-cluster is the minimum energy tree spanning all vertices
of
the IBP clusters}.  This can be seen in Fig.
\ref{fig2}, where the IBP and IBPO clusters are shown
simultaneously.

Since  the IBPO cluster forms a tree,
removing any bond of the IBPO cluster  breaks the cluster in
two subclusters. This is not true for the IBP model, where by
cutting
any trapped bond one does not break the cluster (Fig. \ref{fig2}).
 Since the cluster generated by the IBPO model is a tree,
it has no loops. The fact the IBPO and IBP share the same
scaling exponents shows that loopless IBP (which is the IBPO     
model)
belongs to the same universality class as IBP, or        ordinary
percolation. Loopless percolation       has been studied in great
detail \cite{tzs89},    and there is {\it numerical} evidence  
that
removing loops does not change the universality class   of the
percolation model. However, to my knowledge  the         IBPO model
is the first
percolation  model generating       loopless percolation clusters,
for
which the fact that the loopless model     belongs
to the same universality class
as ordinary (invasion) percolation can be proven exactly.

{\it (e) Dimension dependence---} The proof of (a)-(d) does not
assume
anything specific about the structure of the lattice.  Indeed,
Prim's
theorem applies for an arbitrary weighted graph.  Since any regular
lattice, in any dimension, is a special case of a random graph, the
above results are independent of the nature and dimension of the
lattice, proving (e).

{\it Complexity of the IBPO model---} The number of spanning trees
on a
regular lattice is much smaller that on a fully connected graph,
but
still increases exponentially with $n$.  But the number of
computations needed in the simulation of the invasion processes, or
the {\it complexity of the IBPO algorithm}, is algebraic in $n$
\cite{algo}.  The most time consuming operation is finding at every
time step the bond with the smallest weight eligible for growth.
However, since $N^b_{IBPO} \le N^b_{IBP}$, the IBPO model requires
equal
or less time to run on an arbitrary computer.  Fig. \ref{fig3}
shows
the  number of trapped bonds with time $(N^b_{IBP}-N^b_{IBPO})$.
 Since
the two models belong to the same universality class, using the
IBPO
model for studying the scaling properties of IBP or ordinary
percolation has considerable computational advantages.

        In conclusion, I introduced a new bond invasion percolation
model that belongs to the same universality class as IBP without
trapping, or ordinary percolation.  The cluster generated by the
IBPO
model form the smallest energy tree spanning the IBP cluster. 
Exact
enumeration, which is the only alternative solution to this global
optimization problem, diverges exponentially with the number of
vertices in the system.  This is the first model, to my knowledge,
that through a step-by step optimization process finds the global
minima of the system.

The global optimization problem, to which IBP is shown to be
equivalent, connects to an another class of problems in statistical
mechanics: that of understanding the zero temperature properties of
various spatially extended random systems.  Since the low
temperature
behavior is dominated by configurations with the smallest energy,
such
problems involve finding the minima of certain functions, most
often
of a Hamiltonian.  Problems in physics that regularly deal with
such
minimalization procedures range from directed polymers to spin
glasses \cite{cieplak}, or interface motion in disordered media
\cite{alb}. The IBPO model
provides the minimal energy cluster, implicitly solving a generic
problem for a particular random system whose only other solution is
exact enumeration.

 I have benefited from enlightening discussions with A. Aharony,
 J. Feder, G. Grinstein, S. Havlin, J. Toner, J. T\o ssang, and
Y.Tu.

\begin{figure}
\caption{Definition of the IBP and IBPO models. The figure
shows a portion of a two dimensional square lattice with the
numbers
on the bonds corresponding to the weights $p_{ij}$. Starting the
invasion process from the vertex $A$, both the IBP and IBPO models
select the smallest bonds in the order indicated by the the arrows.
After the third time step the invasion process reaches the vertex
$B$.
The smallest bond is the one marked with a dotted line, connecting
two
already {\it occupied} vertices, $A$ and $B$.  Such a bond is a
{\it
trapped} bond.  The IBP model next  occupies this trapped bond,
without adding any new vertex to the cluster.  However, the trapped
bond is not eligible for growth in the IBPO model, thus the bond
chosen
next by the IBPO model is the one with $p=0.4$.  Observe that in
the
next step the IBP model would choose exactly the same bond. 
Defining
time as the number of occupied {\it vertices}, the two models
occupy
the same vertices in exactly the same order.}
\label{fig1}
\end{figure}

\begin{figure}
\caption{A particular
realization of the IBP and IBPO clusters invading  a square lattice
simultaneously. The solid bonds form the IBPO
cluster. The solid and the dotted (trapped)  bonds together form
the IBP
cluster. Note that the IBPO cluster forms a tree (loopless
cluster),
while every trapped bond leads to a loop on the cluster.}
\label{fig2}
\end{figure}

\begin{figure}
\caption{The difference between the number of bonds occupied by the
IBP and IBPO clusters invading a two dimensional square
lattice with the same $p_{ij}$
configuration.  The horizontal axis corresponds to time, or
$N^b_{IBPO}$, according to the  definition of time used in the
paper,
while the vertical is $(N^b_{IBP}-N^b_{IBPO})$. Note, that since
the difference between the IBP and IBPO models comes in the trapped
bonds,
 $(N^b_{IBP}-N^b_{IBPO})$ coincides  with
 the number of trapped bonds generated by
the IBP  model. The four curves correspond to simulations on 
systems
with size $L \times L$,
where $L=20,50,100,200$. An average over 100 runs was taken for
each curve.
The simulations where stopped after all {\it
vertices} have been occupied. The asymptotic scaling
of the curves suggest a linear behavior.}
\label{fig3}
\end{figure}

\end{multicols}


\begin{references}

\bibitem[*]{byline} Permanent address: Department of Physics,
University of Notre Dame, Notre Dame, IN 46556, Email: alb@nd.edu


\bibitem{wong}
For a recent review see P-z. Wong, MRS Bulletin {\bf 19} (No 5), 
32  (1994); M. Sahimi, {\it Flow and transport in porous media and
fractured rock}  (Weinheim, New York,1995).

\bibitem{Wilkinson83}
D. Wilkinson, and J.F. Willemsen, J.Phys. A {\bf 16}, 3365 (1983);
R. Lenormand, and S. Bories, C.R. Acad. Sci. Paris {\bf 291B}, 279
(1980); R. Chandler, J. Koplik, K. Lerman, and J.F. Willemsen,  J.
Fluid. Mech. {\bf 119}, 249 (1982).


\bibitem{Vicsek92}
T. Vicsek,  {\it Fractal Growth Phenomena} (World
Scientific, Singapore, 1992); J. Feder, {\it Fractals} (Plenum, New
York, 1988);
A. Bunde, and S. Havlin, {\it Fractals and Disordered Systems}
(Springer Verlag, Berlin, 1991).


\bibitem{Chris75}
N. Christofides, {\it Graph theory: An algorithmic approach}
(Academic
Press, London, 1975).


\bibitem{Cayley1874}
A. Cayley, Philosophical Magazine {\bf 67}, 444 (1874).

\bibitem{Prim57}
R.C. Prim, The Bell Syst. Tech. Jl.  {\bf 36} 1389 (1957).

\bibitem{proof}
The detailed proof that the cluster generated by Prim's algorithm
has
the smallest energy is given in \cite{Chris75} and \cite{Prim57}.
For
a short outline of the proof, tailored to the IBPO model, consider
a
graph of $n$ vertices, and assume that we are at the last step of
the
invasion process. The IBPO cluster, $C_{n-1}$, has the smallest
energy
of all trees connecting the selected $n-1$ vertices.  In the next
step
we connect the last vertex, denoted by ${\cal A}$, to the tree,
generating the cluster $C_n$.  This is done by selecting and
occupying
the smallest energy bond connecting ${\cal A}$ to the cluster
$C_{n-1}$.  The proof proceeds by {\it reduction ad absurdum}:
assume
that the obtained cluster $C_n$ does not have the smallest energy,
i.e.  exists a cluster $C_n'$ that has an energy $E(C'_n)<E(C_n)$.
However, this means that $C'_n-{\cal A}$ has smaller energy than
$C_{n-1}$, contradicting the hypothesis that $C_{n-1}$ is the tree
with the smallest energy spanning the $n-1$ vertices. Thus $C_n$
has
to be the smallest energy cluster existing in the system,
concluding
the proof of (a).

\bibitem{trap} The trapping rule used in the IBPO model does not
isolate complete clusters, but only bonds that have both ends
occupied.  Note that the investigated IBP model is the so called IBP
{\it without trapping} \cite{Wilkinson83}.  Similar trapping rules
have been considered by M. Blunt {\it et al.} [M.  Blunt, M.J. King,
and H. Scher, Phys. Rev. A {\bf 46}, 7680 (1992)].


\bibitem{lung}
Opening
mechanisms similar to the one described by the IBPO model  are
known to take place
in the lung during inflation, as it was demonstrated experimentally
by
Suki {\it et al.} \cite{Suki94}.

\bibitem{Suki94}
B. Suki {\it et al.}, Nature {\bf 368}, 615 (1994).

\bibitem{note-bonds}
Note that occupying all vertices does  not imply  that all
bonds have been occupied as well.


\bibitem{unique}
A formal proof of the uniqueness of the shortest spanning tree is
given by
by J.B Kruskal [Proc. Amer. Math. Soc. {\bf 7}, 48 (1956)].


\bibitem{vertex}
 The IBP model selects and occupies at any time step the
smallest of the empty bonds connected to occupied vertices. 
However,
if the selected bond is trapped, occupying it does not occupy any
{
\it new vertex}, i.e. it occupies a bond adding it to the
$B$-cluster,
but the $V$-cluster remains unchanged.  A new vertex is added only
when the selected bond is not trapped (Fig. \ref{fig1}).  By
definition, in the IBPO model only bonds that are not trapped are
eligible for growth, which at every time step are identical to the
IBP
non-trapped bonds. 

\bibitem{univ}
In the following I  argue that   the  IBP and an IBPO cluster,
that
share the
same vertices, must have the same scaling properties, and thus
must belong to the same universality class.  Take a  cluster of
fractal dimension $D_f$ 
generated by  the  invasion bond  percolation algorithm and 
replace every bond with the vertices  connected by the  bonds.
  The performed  {\it local} operation
is not observable  if  the system is viewed at 
length scales  larger than two bond length,
thus   going from
bonds to vertices can 
not affect the  scaling properties of the
cluster.  For example, if we  measure the  fractal
dimension of the  cluster, the differences between the B
and
the V-clusters  come on length scales smaller than two
lattice spacings, i.e. any method that is  investigating the
fractal (large scale) 
properties of the cluster will not  see any difference.  Thus the
 vertex cluster and the bond cluster  belong to the same
universality class. 
Since at any time step the IBP and IBPO models share the
same vertices,  we  have proven that IBP and IBPO models 
belong
to the same universality class. 


\bibitem{Fruenberg}
 L. Furuberg, J. Feder, A. Aharony, and T. J\o ssang, Phys. Rev. 
Lett.
{\bf 61}, 2117--2120 (1988).

\bibitem{tzs89}
F. Tzschichholz, A. Bunde, and S. Havlin, Phys. Rev A {\bf 39},
5470 (1989).

\bibitem{algo}
The execution time of the shortest published program for
generating the minimal  spanning tree using Prim's
algorithm increases as $n^2$ [V. Kevin and M. Whitney, Comm. of ACM
{\bf 15},
273 (1972)].

\bibitem{cieplak} M. Cieplak, A. Maritan, and J.R. Banavar, Phys.
Rev. Lett. {\bf 72}, 2320 (1994);  M. Cieplak, A. Maritan, and J.R.
Banavar (preprint).

\bibitem{alb} A.-L. Barab\'asi and H.E. Stanley, {\it Fractal
Concepts in Surface Growth} (Cambridge University Press,
Cambridge, 1995).

\end{references}
\end{document}